\documentclass[11pt(*)]{JHEP3}
\usepackage{amsmath}
\usepackage{graphicx}
\usepackage{amssymb}
\usepackage{amsthm}
\usepackage{latexsym}
\usepackage{mathrsfs}
\usepackage{epsfig,multicol,bbm}

\newcommand\fverb{\setbox\fverbbox=\hbox\bgroup\verb}
\newcommand\fverbdo{\egroup\medskip\noindent%
            \fbox{\unhbox\fverbbox}\ }
\newcommand\fverbit{\egroup\item[\fbox{\unhbox\fverbbox}]}
\newbox\fverbbox

\title{One-loop effective brane action}

\author{Tae-Hun Lee\\
  Department of Physics, Purdue University,\\
525 Northwestern Avenue, West Lafayette, IN 47907-2036, U.S.A.\\
    E-mail: \email{lee109@physics.purdue.edu}}

\received{\today}
\accepted{\today}
\preprint{\hepth{arXiv:0712.0601}}  

\abstract{The one-loop effective action for a $p$ brane embedded in
a $D=p+2$ Minkowski spacetime in the static gauge is calculated.
Rescaling the quantum fluctuation by $\sqrt{-g_0}$ evaluated on the background brane leads to the one-loop effective
action expressed only in terms of infrared and ultraviolet
divergent geometric scalars. After the infrared divergences are absorbed into the quantum fluctuation, there remains the finite number of ultraviolet divergences. This implies that the $D=p+2$ Poincar\'{e} symmetry and the $D=p+1$ general coordinate invariance
are preserved in one-loop order. }


\dedicated{}

\begin{document}

\section{Introduction}\label{Intro}
Although quantization of a bosonic string allows only a special
dimension, $D=26$ \cite{Polyakov}, brane world scenarios in other
dimensions are still being actively studied in the context of long wavelength brane oscillation effective actions \cite{Clark1,Clark2}. On the other hand, there
could be many possible classically equivalent descriptions of branes which may have different quantum
aspects \cite{Tseytlin}. For example, the Polyakov action \cite{Polyakov} is often a useful choice because
of its rich insight to the physics and convenience due to its covariant nature. Despite the presence of the square root the Nambu-Goto action \cite{Nambu-Goto} is still widely used for fundamental theories as well as description of specific models, for example, numerical analysis of gauge interactions, especially, the strong interaction \cite{Panero}. Especially, in this paper, to see its effective quantum structures of branes in various dimensions the Nambu-Goto action is used to describe a brane embedded in a certain target dimension.\\
The Nambu-Goto action is an invariant world volume element constructed from
the induced metric, so it has the higher dimensional Poincar\'{e}
symmetry as well as the reparametrization invariance. Because of the
reparametrization invariance symmetry the gauge fixing is required. In general, the gauge fixing breaks both of the symmetries. However, the higher dimensional Poincar\'{e}
symmetry can be realized by the nonlinear transformation among parameters and the fields \cite{Clark1,Clark2}. It is a question whether this nonlinear symmetry is preserved in every order of loops. The purpose of this paper is
to calculate the one-loop effective action for a $p$ brane embedded
in a $D=p+2$ Minkowski spacetime in the static gauge \cite{Clark1,Clark2}, focusing on the symmetries.\\
Specifically, the action for a $p$ brane embedded in a $D=p+2$ Minkowski spacetime can be described by a single
scalar field in the static gauge, where the ghost contribution is
absent as it is in the axial gauge in non-Abelian gauge theory. In a
naive expansion of the action around the classical field, the term
corresponding to one-loop order appears as if it were a scalar field
action in the gravitational field, denoted by the 0 subscript, as the classical non-tensorial object $g^{-2/(p-1)}_0g_{0mn}$, where $g_0\equiv-\det g_{0mn}$. Hence, the classical symmetries of the
effective action are not obviously preserved. One methodology to attempt to maintain the symmetry structure is to return to the original action and expand the
action about classical fields without fixing the gauge. The Faddeev-Popov gauge
fixing and ghost terms must be added to the action. It turns out that it is difficult to
integrate the generating functional to get the determinant of the
double index non-symmetric metric where the $D=p+2$ and $D=p+1$
Lorentz indices are entangled. The geodesic expansion \cite{Honerkamp,Brown}
could be another possible way though it was not considered in this paper.\\
This paper shows that the effective action in one-loop order can be
expressed in terms of geometric scalars by rescaling the
quantum fluctuation by its own classical Lagrangian density
$\sqrt{-g_0}$. The path integral measure is not changed by
this local scale change in the dimensional regularization, contrasted to the
canonical, Hamiltonian formalism \cite{Fujikawa}. Therefore, a ghost action
due to the Jacobian factor from the local scale change does not appear.
But the origin of this scale transformation and the complete understanding of the consequent divergent
structure still remain a question to be addressed.
\section{The Nambu-Goto action in the static gauge}\label{Nambu-Goto}
The symmetries of the brane action are made manifest in one-loop in this section. It will be shown that this can be done by
rescaling the quantum fluctuation.
The action for a $p$ brane embedded in a $D=p+2$ Minkowski spacetime is given as the Nambu-Goto action \cite{Nambu-Goto}.\\
\begin{equation}
S=-\sigma\int d^{p+1}x \sqrt{-\det[\eta_{\mu\nu}\partial_m
X^\mu(x)\partial_n X^\nu(x)]},\label{eq:S}
\end{equation}
where $\mu,\nu=0,1,2,\cdots,p+1$ and $m,n=0,1,2, \cdots, p$. It is well known that it has the
$D=p+2$ Poincar\'{e} symmetry of the bulk as well as the reparametrization invariance,
or equivalently, the gauge symmetry on the scalar fields $X^\mu$s,
i.e., $X^\mu\rightarrow X^\mu+\omega^a\partial_aX^\mu$ \cite{Hooft}. The
static gauge allows one to fix the gauge by simply choosing the
parameters $x^m$ equal to the fields $X^m$. In the path integral method, the
gauge fixing condition $f^m=X^m-x^m$ yields the Jacobian in the
functional integration measure, $\det[\frac{\delta
f^m(x)}{\delta\omega^n(y)}]=\det[\delta(x-y)\partial_nX^m]$, which
becomes independent of $X^m$ when combined with $\delta[f^m-\chi^m]$, where $\chi^m$ is an arbitrary function to be integrated out in the path integral.
Therefore, there is no ghost contribution in the static gauge as in
the axial gauge in non-Abelian gauge theory. With the only remaining
degree of freedom the $(p+2)$th coordinate, $\phi\equiv X^{p+1}$,
the Lagrangian density becomes \cite{Clark1,Clark2}
\begin{equation}
\mathscr{L}=-\sigma\sqrt{1-\partial_m\phi\partial^m\phi}=-\sigma\sqrt{-\det g_{mn}}\equiv -\sigma g^{\frac{1}{2}},
\label{eq:L}
\end{equation}
where $\partial^m\equiv\eta^{mn}\partial_n$,
$g_{mn}=\eta_{mn}-\partial_m\phi\partial_n\phi$ and its inverse
$g^{mn}=\eta^{mn}+g^{-1}\partial^m\phi\partial^n\phi$. The effective action $\Gamma$ is defined by,
\begin{equation}
e^{\frac{i}{\hbar}\Gamma[\phi_0,J]}=\int\mathscr{D}\phi\exp \frac{i}{\hbar}\int
d^{p+1}x[\mathscr{L}(\phi)+(\phi -\phi_0)J]=\int\mathscr{D}\varphi\exp
\frac{i}{\hbar}\int d^{p+1}x[\mathscr{L}(\varphi+\phi_0)+\varphi J],\label{eq:eGamma}
\end{equation}
Expanding the shifted action about a classical field $\phi_0$ and keeping it up to the second order in the fluctuation $\varphi$ as $\phi=\varphi+\phi_0$,
\begin{equation}
\begin{array}{ccl}
\int d^{p+1}x\mathscr{L}(\varphi+\phi_0) &=&-\sigma\int
d^{p+1}xg^{\frac{1}{2}}_0+\sigma\int
d^{p+1}xg^{-\frac{1}{2}}_0\partial^m\phi_0\partial_m\varphi\\
&&+\frac{1}{2}\sigma\int d^{p+1}xg^{-\frac{1}{2}}_0g^{mn}_0\partial_m\varphi\partial_n\varphi+\cdots\\
&=&\sigma\int d^{p+1}xg^{\frac{1}{2}}_0(-1
+\frac{1}{2}g^{-1}_0g^{mn}_0\partial_m\varphi\partial_n\varphi+\cdots),
  \end{array}\label{eq:Lphi0}
\end{equation} where the linear term has been eliminated by the classical equation of motion
$\partial^m(g^{-\frac{1}{2}}_0\partial_m\phi_0)=0$. The
quadratic term corresponding to one-loop order looks like the scalar
field action in curved spacetime with the metric $\tilde{g}_{0mn}$. By taking a determinant of the background part
\begin{equation}
g^{-\frac{1}{2}}_0g^{mn}_0=\tilde{g}^{\frac{1}{2}}_0\tilde{g}^{mn}_0,
\end{equation}
the metric for such a gravitational field $\tilde{g}_{0mn}$ can be identified with
\begin{equation}
\tilde{g}_{0mn}=g^{-2/(p-1)}_0g_{0mn}.
\end{equation}
However, $g^{s}_0g_{0mn}(s\neq0)$ does not transform like a tensor under the
$D=p+2$ Poincar\'{e} symmetry \cite{Clark1,Clark2} (For the details see Eqs.(\ref{eq:Deltag}, \ref{eq:g-1gmn1}, \ref{eq:g-1gmn2}) in \textbf{Sec. \ref{Poincare})}. In the end, the effective action would not be expected
to be manifestly invariant under these symmetries. In order to allow for a manifestly invariant action a classical field dependent background rescaling of the fluctuation is utilized.
\begin{equation}
\varphi\rightarrow \alpha\varphi,\label{eq:scale}
\end{equation}
where $\alpha=\alpha(\phi_0)$ is a local function of $\phi_0$. This scale
transformation changes the functional measure by an infinite
constant or $\delta(0)\times \textrm{constant}$, in which
$\delta(0)$ vanishes in the dimensional regularization.
\begin{equation}
\mathscr{D}\varphi\rightarrow\mathscr{D}\varphi\det[\alpha]
=\mathscr{D}\varphi e^{\delta(0)\int d^{p+1}x\ln
\alpha}=\mathscr{D}\varphi.\label{eq:measure}
\end{equation}
The scale change produces $\varphi^2$ term in the action.
\begin{equation}
\begin{array}{ccl}
\int
d^{p+1}xg^{-\frac{1}{2}}_0g^{mn}_0\partial_m(\alpha\varphi)\partial_n(\alpha\varphi)
&=&\int
d^{p+1}xg^{-\frac{1}{2}}_0g^{mn}_0\alpha^2\partial_m\varphi\partial_n\varphi\\
&&+\int
d^{p+1}x[g^{-\frac{1}{2}}_0g^{mn}_0\partial_m\alpha\partial_n\alpha-\partial_n(g^{-\frac{1}{2}}_0g^{mn}_0\frac{1}{2}\partial_m\alpha^2)]\varphi^2.
\end{array}\label{eq:2ndorder}
\end{equation}
If $\alpha$ chosen to be $g^{\frac{1}{2}}_0$, the first
term is obviously in a manifestly invariant form. Moreover, as explicitly shown in the
\textbf{Appx. \ref{Scale}}, using the equation of motion the background parts in the second term reduce to
\begin{equation}
g^{-\frac{1}{2}}_0g^{mn}_0\partial_mg^{\frac{1}{2}}_0\partial_ng^{\frac{1}{2}}_0-\partial_n[g^{-\frac{1}{2}}_0g^{mn}_0\frac{1}{2}\partial_m(g^{\frac{1}{2}}_0)^2]
= g^{\frac{1}{2}}_0R_0,\label{eq:gR}
\end{equation} where $R_0$ is the Ricci scalar
based on $g_{0mn}$. Thus, it is obtained that $D=p+2$ Poincar\'{e} invariant world volume action for the brane is secured in one-loop order. (The detailed explanation is provided in the next section.) As a result of the rescaling, the effective action in one-loop at $J=0$ is
\begin{equation}
\Gamma[\phi_0]_{\textrm{one-loop}}=-i\hbar\ln\int\mathscr{D}\varphi\exp
\frac{i}{\hbar}\sigma\int d^{p+1}x\frac{1}{2}g^{\frac{1}{2}}_0(
g^{mn}_0\partial_m\varphi\partial_n\varphi+R_0\varphi^2).\label{eq:Gamma}
\end{equation} It can be immediately recognizable that except for $\sigma$ this is the same as the generating function
of the connected Green function for a massless scalar field in
background gravity, $W[0]$ in $D=p+1$ with $\xi=-1$ \cite{Birrell-Davies}.
\begin{equation}
W[0]=-i\hbar\ln\int\mathscr{D}\phi\exp\frac{i}{\hbar}\int
d^{p+1}x\frac{1}{2}\sqrt{-g}[g^{mn}\partial_m\phi\partial_n\phi-(m^2+\xi
R)\phi^2].\label{eq:W}
\end{equation}
\subsection{$D=p+2$ Poincar\'{e} symmetry}\label{Poincare}
This section is intended to explicitly show that the metric $g_{0mn}$
and its inverse $g^{mn}_0$ can be treated as a tensor on the $D=p+2$
Poincar\'{e} symmetry and consequently any geometrical tensor made
of these
metrics and derivatives is also such a tensor. (Note that the indices in $(x_m,b_m,v_m)$ are raised by $\eta^{mn}$, for a shorthand notation 0 subscript is dropped and the following notations are used. $\partial_m\phi\equiv v_m$, $b\cdot v\equiv \eta^{mn}b_mv_n$, $v^2\equiv v_m v^m$, etc.)\\
In the static gauge the $D=p+2$ Poincar\'{e} symmetry is
realized on the coordinates and field \cite{Clark1,Clark2}, as
\begin{equation}
\begin{array}{ccl}
x^{\prime m}&=&x^m+a^m-\phi b^m+\epsilon^{mnr}\alpha_nx_r,\\
\Delta\phi&=&z-b_mx^m,
\end{array}\label{eq:Trans1}
\end{equation}
where $\Delta$ is a total variation and $z$ and $a^m$ are a broken and an unbroken infinitesimal translational transformation
parameter, respectively while $b_m$ and $\alpha_n$ are a broken and an unbroken
infinitesimal Lorentz transformation parameter, respectively. Since the action still has the
$D=p+1$ unbroken Poincar\'{e} symmetry, the only parts to be
considered in Eq.(\ref{eq:Trans1}) are the $D=p+2$ higher dimensional broken symmetry transformations. Note that the constant $z$ can be ignored since the action depends on only derivatives of $\phi$.
\begin{equation}
\begin{array}{ccl}
x^{\prime m}&=&x^m-\phi b^m,\\
\Delta\phi&=&-b_mx^m.
\end{array}\label{eq:Trans2}
\end{equation}
 If $g_{mn}$ and $g^{mn}$ transform under the given
transformations as tensors $
g^\prime_{mn}=\frac{\partial x^p}{\partial x^{\prime
m}}\frac{\partial x^q}{\partial x^{\prime n}}g_{pq}$ and $ g^{\prime
mn}=\frac{\partial x^{\prime m}}{\partial x^p}\frac{\partial
x^{\prime n}}{\partial x^q}g^{pq} $, technically they can be treated
as a tensor.
\begin{equation}
\begin{array}{ccl}
g^\prime_{mn}&=&\frac{\partial x^p}{\partial x^{\prime m}}\frac{\partial x^q}{\partial x^{\prime n}}g_{pq}\\
&=&(\delta^p_n+b^pv_n)(\delta^q_m+b^qv_m)g_{pq}\\
&=&\eta_{mn}-v_mv_n-b_mv_n-b_nv_m+2(b\cdot v)v_mv_n\\
&=&g_{mn}-b_mv_n-b_nv_m+2(b\cdot v)v_mv_n,\\
g^{\prime mn}&=&\frac{\partial x^{\prime m}}{\partial x^p}\frac{\partial x^{\prime n}}{\partial x^q}g^{pq}\\
&=&(\delta^n_p-b^nv_p)(\delta^m_q-b^mv_q)g^{pq}\\
&=&\eta^{mn}-b^nv^m-b^mv^n+g^{-1}v^mv^n-g^{-1}v^mb^nv^2-g^{-1}v^nb^mv^2\\
&=&g^{mn}-g^{-1}b^mv^n-g^{-1}b^nv^m.
\end{array}\label{eq:gmn1}
\end{equation}
 By using the transformations
$v_m^\prime=v_m+v_m(b\cdot v)-b_m $ and $g^\prime=g+2g(b\cdot v)$ and their definitions $g_{mn}=\eta_{mn}-v_mv_n$ and $g^{mn}=\eta^{mn}+g^{-1}v^mv^n$, it
can be checked that they indeed follow the
same transformations.
\begin{equation}
\begin{array}{ccl}
g^\prime_{mn}&=&g_{mn}+\delta[\eta^\prime_{mn}]+\delta[v_m]v_n+v_m\delta[v_n]\\
&=&g_{mn}+[v_m(b\cdot v)-b_m]v_n+v_m[v_n(b\cdot v)-b_n]\\
&=&g_{mn}-b_mv_n-b_nv_m+2v_mv_n(b\cdot v),\\
g^{\prime mn}&=&g^{mn}+(-g^{-2})(2gb\cdot v)v^mv^n+g^{-1}\delta[v^m]v^n+g^{-1}v^m\delta[v^n]\\
&=&g^{mn}-2g^{-1}b\cdot vv^mv^n+g^{-1}(2v^mv^nb\cdot v-b^mv^n-b^nv^m)\\
&=&g^{mn}-g^{-1}b^mv^n-g^{-1}b^nv^m.
\end{array}\label{eq:gmn2}
\end{equation} Therefore, any geometric tensor based on these metric tensors
transforms like a tensor under the given transformations. Based on these transformations, it can be seen that $g^sg_{mn}(s\neq0)$ does not transform as a tensor since $g$ is not a scalar. Explicitly,
\begin{equation}
\Delta g=-2v^m\Delta[v_m]=-2(v_mb\cdot v-b_m)v^m=2g(b\cdot v)\label{eq:Deltag}
\end{equation}
and thus
\begin{equation}
\begin{array}{ccl}
[g^sg_{mn}]^\prime&=&g^sg_{mn}+sg^{s-1}\Delta gg_{mn}+g^{s}\Delta g_{mn}\\
&=&g^sg_{mn}+sg^{s-1}2g(b\cdot v)g_{mn}+g^s[-b_mv_n-b_nv_n+2v_mv_n(b\cdot v)]\\
&=&g^s[g_{mn}(1-2s(b\cdot v))-b_mv_n-b_nv_n+2v_mv_n(b\cdot v)].
\end{array}\label{eq:g-1gmn1}
\end{equation}
This is not a tensor transformation, that is,
\begin{equation}
[g^sg_{mn}]^\prime\neq \frac{\partial x^p}{\partial x^{\prime
m}}\frac{\partial x^q}{\partial x^{\prime n}}g^sg_{pq}=g^s[g_{mn}-b_mv_n-b_nv_m+2(b\cdot v)v_mv_n].
\label{eq:g-1gmn2}
\end{equation}
Hence, $g^{-2/(p-1)}_0g_{0mn}$ in the action is not a tensor.
\subsection{Scalar field in background gravity}\label{Scalar field}
In the previous sections, it has been seen that the one-loop
effective action is given as $W[J=0]$ in a scalar field theory in
background gravity. Here, the result of calculation of $W[J=0]$ is introduced from [Birrell \& Davies, (1982)]\cite{Birrell-Davies}.
Suppose a scalar field is in a background curved spacetime described by the Lagrangian density\\
\begin{equation}
\mathscr{L}=\frac{1}{2}\sqrt{-g}[g^{mn}\partial_m\phi\partial_n\phi-(m^2+\xi
R)\phi^2],\label{eq:Lscalar}
\end{equation}
 where $\xi$ is a numerical factor, $m$ is
the mass of the field $\phi$ and $R$ is the Ricci scalar. The generating
functional $Z$ and the generating function of the connected Green
functions $W$ in $D=p+1$ are
\begin{equation}
Z[J]=\int\mathscr{D}\phi\exp[\frac{i}{\hbar}\int d^{p+1}x(\mathscr{L}+J\phi)],~~
W=-i\hbar\ln Z[0].\label{eq:ZW}
\end{equation}
In the dimensional regularization, $W$ is given in Eq.(6.41), p.159, Birrell \&
Davies, (1982) \cite{Birrell-Davies}.
\begin{equation}
\begin{array}{ccl}
W&=&\lim\limits_{n\to p+1}\hbar\int d^nx\sqrt{-g}\frac{1}{2}(4\pi)^{-n/2}\sum\limits^\infty_{j=0}
a_j(x)\int^\infty_0(is)^{j-1-n/2}e^{-im^2s}ids\\
&=&\lim\limits_{n\to p+1}\hbar\int d^nx
\sqrt{-g}\frac{1}{2}(4\pi)^{-n/2}\sum\limits^\infty_{j=0}a_j(x)(m^2)^{n/2-j}\Gamma(j-n/2),
\end{array}\label{eq:Ws}
\end{equation}
where
\begin{equation}
\begin{array}{ccl}
&&a_0(x)=1,\\
&&a_1(x)=(\frac{1}{6}-\xi)R,\\
&&a_2(x)=\frac{1}{180}R_{abcd}R^{abcd}-\frac{1}{180}R^{ab}R_{ab}-\frac{1}{6}(\frac{1}{5}-\xi)\square
R+\frac{1}{2}(\frac{1}{6}-\xi)^2R^2,\\
&&\cdots.
\end{array}\label{eq:a012}
\end{equation}
It is necessary to recognize that $a_j(x)$ is a $2j-$derivative geometric scalar with respect to the metric. This implies that the net remaining number of $g^{mn}$ after contraction with $g_{mn}$ in $a_j(x)$ is $j$.
\subsection{One-loop effective action}\label{One-loop}
Now the above result Eq.(\ref{eq:Ws}) can be identified with the one-loop effective brane action when the mass parameter is sent to zero. This limit produces an infinite series of divergences due to $(m^2)^{n/2-j}$ when $n/2-j<0$. Regardless of this limit, there is another source of divergence from $\Gamma(j-n/2)$ when $n/2-j\geq0$. One can think that in Eq.(\ref{eq:Ws}) $e^{-im^2s}$ is introduced just for regulating the first divergences in the DeWitt-Schwinger representation of a massless scalar field action. The origin of these divergences comes from $a_j(x)\int d^nk e^{-iky}(k^2-m^2)^{-j-1}$ in the propagator expansion, together with the extra power factor $m^2$ from the final effective action $W[0]=-\frac{1}{2}i\mbox{tr}[\ln(-G_F)]$ when $n/2-j<0$ \cite{Birrell-Davies}, for which zero momentum mode gives a divergence in the massless limit. Similarly, the divergences in $\Gamma(j-n/2)$ originate from the other limit $k\rightarrow\infty$ when $n/2-j\geq 0$. In this reason, they  can be called infrared and ultraviolet divergences, respectively. (For the detailed power counting of the mass parameter, see Eq.(3.130), p.74 for the propagator expression and also Eq.(6.34), p157 in Birrell \& Davies, (1982))\cite{Birrell-Davies}.\\
 The infrared divergences $(m^2)^{n/2-j}$ in Eq.(\ref{eq:Ws}) seem to be somewhat fictitious. A constant scale change of $\varphi$ causes only adding number in the effective action Eq.(\ref{eq:Gamma}). It is equivalent to the corresponding scale change in $g_{mn}$. Therefore, if this scale is properly chosen to cancel $(m^2)^{n/2-j}$ factor, i.e.,
 \begin{equation}
  \varphi\rightarrow(m^2)^{-n/4+1/2}\varphi,
 \end{equation}
 equivalently,
 \begin{equation}
 g_{0mn}\rightarrow (m^2)^{-1}g_{0mn}
 \end{equation}
 and hence
 \begin{equation}
 a_j\rightarrow (m^2)^ja_j.
 \end{equation}
 They bring the scale $(m^2)^{-n/2+j}$ in the one-loop effective action Eq.(\ref{eq:Gamma}). As a result, only ultraviolet divergences remain in the one-loop effective action.
 \begin{equation}
 \Gamma[\phi_0]_{\mbox{one-loop}}=\lim\limits_{n\to p+1}\frac{\hbar}{2}(4\pi)^{-n/2}\sigma^{n/(n-2)}\int d^nx
g^{\frac{1}{2}}_0\sum\limits^\infty_{j=0}\sigma^{-2j/(n-2)} a_{j0}(x)\Gamma(j-n/2)
 \end{equation}
 The terms with $j>n/2$ in the effective action are finite. Therefore, there remains only the finite number of ultraviolet divergences, which can be renormalized by adding counter terms. In addition,
these results can be always simplified further by using the constraint Eq.(\ref{eq:IdentityA}) from the equation of motion. (For the derivation, refer to\textbf{ Appx. \ref{Curvature})}
\begin{equation}
R^{abcd}_0R_{0abcd}=2(R^2_0-R^{ab}_0R_{0ab}).\label{eq:Identity}
\end{equation}
where note that in $n=2$ case this relation is just a general identity, not a constraint.
\subsection{Effective correction to Einstein's field equation}\label{Effective}
Since the quantum corrections to the classical brane action have the $D=p+1$ general coordinate invariance as well as the $D=p+2$ Poincar\'{e} symmetry, they can be effectively treated as
gravitational interaction with the metric $g_{0mn}$. If the higher order extra corrections in the effective action can be considered a new effective contribution to a gravitational action, the effective classical field equation can be constructed with modification by these terms. For example, for $p+1=4$ case \cite{Birrell-Davies}, as the infinite terms including $a_0$, $a_1$ and $a_2$ require renormalization of the cosmological constant $\Lambda$, the gravitational constant $G$ from $R$ and new couplings from $\frac{1}{180}R_{abcd}R^{abcd}-\frac{1}{180}R^{ab}R_{ab}-\frac{1}{6}(\frac{1}{5}-\xi)\square R$, respectively, the last higher order contribution effectively modifies the field equation.
\begin{equation}
R_{0mn}-\frac{1}{2}R_0g_{0mn}+\Lambda g_{0mn}+a H^{(1)}_{0mn}+b
H^{(2)}_{0mn}+c H_{0mn}=0,\label{eq:Eeq}
\end{equation}
where $a$, $b$ and $c$ are the corresponding renormalized coefficients after the divergent coefficients are absorbed into their bare couplings and the higher derivative tensors \cite{Birrell-Davies} are
\begin{equation}
\begin{array}{ccl}
H^{(1)}_{0mn}&\equiv& g^{-\frac{1}{2}}_0\frac{\delta}{\delta
g^{mn}_0}\int d^4x g^{\frac{1}{2}}_0R^2_0\\
&=&2R_{0;mn}-2g_{0mn}\square
R_0-\frac{1}{2}g_{0mn}R^2_0+2R_0R_{0mn},\\
H^{(2)}_{0mn}&\equiv& g^{-\frac{1}{2}}_0\frac{\delta}{\delta
g^{mn}_0}\int d^4x
g^{\frac{1}{2}}_0R_{0ab}R^{ab}_0\\
&=&2R^a_{0m;na}-\square
R_{0mn}-\frac{1}{2}g_{0mn}\square
R_0+2R^a_{0m}R_{0an}-\frac{1}{2}g_{0mn}R^{ab}_0R_{0ab},\\
H_{0mn}&\equiv& g^{-\frac{1}{2}}_0\frac{\delta}{\delta g^{mn}_0}\int
d^4xg^{\frac{1}{2}}_0 R_{0abcd}R^{abcd}_0\\
&=&-\frac{1}{2}g_{0mn}R^{abcd}_0R_{0abcd}+2R_{0mabc}R^{abc}_{0n}-4\square
R_{0mn}+2R_{0;mn}-4R_{0ma}R^a_{0n}\\
&&+4R^{ab}_0R_{0ambn},
\end{array}\label{eq:H}
\end{equation}
where note that the term $\square R_0$ in $a_2$ has been ignored because $g^{\frac{1}{2}}_0\square R_0$ is a total derivative and thus $\int d^4x g^{\frac{1}{2}}_0\square R_0=0$ and for simplification $H_{mn}=-H^{(1)}_{mn}+4H^{(2)}_{mn}$
may be applied, obtained from the well-known identity in $D=4$,
$g^{-\frac{1}{2}}\frac{\delta}{\delta g^{mn}}\int
d^4x(R_{abcd}R^{abcd}+R^2-4R_{ab}R^{ab})=0$ \cite{Birrell-Davies}.
 \section{Conclusion}\label{Conclusion}
The action for a $p$ brane embedded in a $D=p+2$ Minkowski spacetime has been examined in one-loop order.
It was found that the special local scale transformation enables the $D=p+1$ general
coordinate invariance and the $D=p+2$ Poincar\'{e} symmetry to be
preserved in classical geometric form in one-loop. These geometric scalars
include ultraviolet divergences and an infinite series of infrared divergences. However, an infinite series of the infrared divergences can be removed by the global scale change.
\appendix
\section{Notations}\label{Notation}
 Greek ($\mu,\nu,\cdots$) and Latin letters ($m, n,
\cdots$) are used for $D=p+2$ and $D=p+1$ Lorentz indices,
respectively. The metric convention $\eta_{mn}=(1,-1,-1,-1,\cdots)$
is used. Also, the following notations are introduced for simplicity.
$$v_m\equiv\partial_m\phi,~ v_{mn}\equiv\partial_m\partial_n\phi ,~ \partial_{mn}\equiv\partial_m\partial_n,~  v^m\equiv \eta^{mn}v_n,~v^2=v_mv^m,~g \equiv \det (-g_{mn}),$$
$$
p_{,\alpha}\equiv\partial_\alpha p,~p^\alpha_{;\beta}\equiv
p^\alpha_{,\beta}+\Gamma^\alpha_{\mu\beta}p^\mu,~p_{\alpha;\beta}\equiv
p_{\alpha,\beta}-\Gamma^\mu_{\alpha\beta}p_\mu.
$$
For instance, $g_{mn}=\eta_{mn}-v_m v_n$ and
$g^{mn}=\eta^{mn}+g^{-1}v^m v^n$.
\section{Curvature tensors and Scale transformation}\label{CurvatureScale}
This section provides explicit derivations of all the necessary geometric scalars necessary for Eq.(\ref{eq:a012}) and Eq.(\ref{eq:gR}) with the metric $g_{mn}=\eta_{mn}-v_mv_n$.
\subsection{Christoffel symbol}\label{Christoffel}
The corresponding Christoffel symbol is
\begin{equation}
\begin{array}{ccl}
\Gamma^a_{bc}&=&\frac{1}{2}g^{am}(g_{mb,c}+g_{mc,b}-g_{bc,m})\\
&=&\frac{1}{2}(\eta^{am}+\frac{1}{g}v^av^m)(-2v_m
v_{bc})\\
&=&-g^{-1}v^a v_{bc}.
\end{array}\label{eq:Chris}
\end{equation}

\subsection{Extrinsic curvature}\label{Extrinsic}
Since all the geometric scalars can be expressed with the extrinsic curvature and it is also a tensor under the higher dimensional Poincar\'{e} symmetry, it may be convenient to introduce it. The extrinsic curvature is defined as follows.
\begin{equation}
K_{ab}=n_{\alpha;\beta}e^\alpha_ae^\beta_b,\label{eq:Kab}
\end{equation}
 where the
unit normal vector
$n_\alpha=\frac{\epsilon\Phi_{,\alpha}}{\sqrt{|g^{\mu\nu}\Phi_{,\mu}\Phi_{,\nu}|}}$
 to the surface $\Phi(X)=0$ and $\epsilon=\pm1$ ($+$ for a timelike surface, $-$ for a spacelike
surface). For a $p$ brane hypersurface embedded in a $D=p+2$
Minkowski spacetime, $\Phi(X)=\phi(x)-X^{p+1}=0$ where $\phi(x)$ is an arbitrary function of $x$,
\begin{equation}
n_a=\frac{\epsilon
v_a}{g^{\frac{1}{2}}},~n_{p+1}=-\frac{\epsilon}{g^{\frac{1}{2}}},~
e^i_a=\delta^i_a,~e^{p+1}_a=v_a.\label{eq:na}
\end{equation}
Using the only nonzero components, $
n_{p+1;m}=-\frac{\epsilon}{2}g^{-\frac{3}{2}}\partial_mv^2$ and
$n_{m;n}=\epsilon
g^{-\frac{1}{2}}v_{mn}+\frac{\epsilon}{2}g^{-\frac{3}{2}}\partial_nv^2v_m
$, the extrinsic curvature is found.
\begin{equation}
K_{mn}=g^{-\frac{1}{2}}v_{mn},\label{eq:Kmn}
\end{equation}
where $\epsilon$ is
chosen to be 1. Note this is a symmetric tensor. When the scalar
$K_{mn}g^{mn}$ is made of the classical field, it vanishes by the equation of motion.
\begin{equation}
K_{0mn}g^{0mn}=
g^{-\frac{1}{2}}_0v_{0mn}(\eta^{mn}+g^{-1}_0v^m_0v^n_0)=g^{-\frac{1}{2}}_0(\partial\cdot
v_0+\frac{1}{2g_0}v_0\cdot\partial v^2_0)=0.\label{eq:Kg}
\end{equation}
\subsection{Curvature tensors}\label{Curvature}
All the curvature tensors are functions of the extrinsic curvature
tensors and the metric.
\begin{equation}
\begin{array}{ccl}
&&R_{abcd}=\frac{1}{g}(v_{ad}v_{bc}-v_{bd}v_{ac})=K_{ad}K_{bc}-K_{bd}K_{ac},\\
&&R^{abcd}=K^{ad}K^{bc}-K^{bd}K^{ac},\\
&&R_{ab}=g^{mn}R_{manb}=g^{mn}(K_{mb}K_{na}-K_{mn}K_{ab})=K_{am}K^m_b-K_{ab}K,\\
&&R^{ab}=g^{am}g^{bn}R_{mn}=g^{am}g^{bn}(K_{mp}K^p_n-K_{mn}K)=K^a_pK^{bp}-K^{ab}K,
\end{array}\label{eq:Riemann1}
\end{equation}
where $K\equiv g^{ab}K_{ab}$, $K^a_b\equiv g^{am}K_{mb}$ and
$K^{ab}\equiv g^{am}g^{bn}K_{mn}$. All the curvature scalars are
simply expressed in terms of the contractions of extrinsic curvature
tensors.
\begin{equation}
\begin{array}{ccl}
&&R=g^{ab}(K_{am}K^m_b-K_{ab}K)=K^a_bK^b_a-K^2,\\
&&\begin{array}{ccl}R^{ab}R_{ab}&=&(K^a_nK^{bn}-K^{ab}K)(K_{am}K^m_b-K_{ab}K)\\
&=&K^a_nK^n_bK^b_mK^m_a-2K^a_mK^m_bK^b_aK+K^a_bK^b_aK^2,
\end{array}\\
&&R^{abcd}R_{abcd}=2[(K^a_bK^b_a)^2-K^a_bK^b_cK^c_dK^d_a].
\end{array}\label{eq:Riemann2}
\end{equation}
With $K_0=0$, the identity can be easily obtained.
\begin{equation}
R^{abcd}_0R_{0abcd}=2(R^2_0-R^{ab}_0R_{0ab}).\label{eq:IdentityA}
\end{equation}
The tensors needed to calculate the four-derivative curvature scalars
are
\begin{equation}
\begin{array}{ccl}
&&K^a_b=(\eta^{am}+\frac{1}{g}v^av^m)g^{-\frac{1}{2}}v_{mb}
=g^{-\frac{1}{2}}v^a_b+v^a\partial_bg^{-\frac{1}{2}}
=\partial_b(g^{-\frac{1}{2}}v^a),\\
&&K^m_bK_{ma}=g^{-\frac{1}{2}}v_{mb}(g^{-\frac{1}{2}}v^m_a+\frac{1}{2}g^{-\frac{3}{2}}v^m\partial_av^2)
=g^{-1}(v_{mb}v^m_a+\frac{1}{4g}\partial_av^2\partial_bv^2),\\
&&K^m_aK^b_m
=g^{-3}[g^2v^m_av^b_m+\frac{g}{4}\partial_av^2\partial^bv^2+\frac{g}{2}v^bv^m_a\partial_mv^2+\frac{1}{4}(v\cdot\partial
v^2)v^b\partial_av^2],\\
&&\begin{array}{ccl} K^m_aK^b_mK^c_b
&=&g^{-\frac{9}{2}}[g^3v^m_av^b_mv^c_b+\frac{g^2}{4}\partial_av^2\partial^bv^2v^c_b+\frac{g^2}{4}\partial^cv^2v^m_a\partial_mv^2+\frac{g}{8}(v\cdot\partial
v^2)\partial^cv^2\partial_av^2]\\
&&+g^{-\frac{9}{2}}[\frac{g^2}{2}v^cv^m_av^b_m\partial_bv^2+\frac{g}{8}v^c\partial_av^2\partial
v^2\cdot\partial v^2+\frac{g}{4}v^c(v\cdot\partial
v^2)v^m_a\partial_mv^2\\
&&+\frac{1}{8}v^c(v\cdot\partial
v^2)^2\partial_av^2].
\end{array}
\end{array}\label{eq:Ktensor}
\end{equation}
Then, the scalars for the four-derivative curvature scalars are constructed from them.
\begin{equation}
\begin{array}{ccl}
&&K^m_aK^a_m=g^{-3}[g^2v^m_av^a_m+\frac{g}{2}\partial
v^2\cdot\partial v^2+\frac{1}{4}(v\cdot\partial v^2)^2],\\
&&K^m_aK^b_mK^a_b
=g^{-\frac{9}{2}}[g^3v^m_av^b_mv^a_b+\frac{3g^2}{4}\partial_av^2\partial^bv^2v^a_b+\frac{3g}{8}(v\cdot\partial
v^2)(\partial v^2\cdot\partial v^2)+\frac{1}{8}(v\cdot\partial
v^2)^3],\\
&&\begin{array}{ccl} K^m_aK^b_mK^c_bK^a_c
&=&g^{-6}[g^4v^m_av^b_mv^c_bv^a_c+g^3\partial_av^2\partial^bv^2v^c_bv^a_c+\frac{g^2}{2}(v\cdot\partial
v^2)\partial^cv^2\partial_av^2v^a_c\\
&&+\frac{g^2}{8}(\partial v^2\cdot\partial v^2)^2
+\frac{g}{4}(v\cdot\partial v^2)^2(\partial v^2\cdot\partial
v^2)+\frac{1}{16}(v\cdot\partial v^2)^4].
\end{array}
\end{array}\label{eq:Kscalar}
\end{equation}
\subsection{Scale transformation}\label{Scale}
~Note that for simplicity 0 subscript denoting the classical field is
omitted here. With $\alpha=g^{\frac{1}{2}}$ and
$\partial_m\alpha=\frac{1}{2}g^{-\frac{1}{2}}\partial_m(-v^2)$. In
Eq.(\ref{eq:2ndorder})
\begin{equation}
\begin{array}{ccl}
&&g^{-\frac{1}{2}}g^{mn}\partial_m g^{\frac{1}{2}}\partial_n
g^{\frac{1}{2}}
-\frac{1}{2}\partial_n(g^{-\frac{1}{2}}g^{mn}\partial_m(g^{\frac{1}{2}})^2)\\
&=&\frac{3}{4}g^{-\frac{3}{2}}\partial v^2\cdot\partial
v^2+g^{-\frac{5}{2}}(v\cdot\partial
v^2)^2+\frac{1}{2}g^{-\frac{3}{2}}\partial\cdot v(v\cdot\partial
v^2) +\frac{1}{2}g^{-\frac{1}{2}}(\partial^2 v^2+g^{-1}v^m
v^n\partial_{mn} v^2).
\end{array}\label{eq:2ndorderA}
\end{equation}
Noticing the identities,
\begin{equation}
\begin{array}{ccl}
&&\partial\cdot v=-\frac{1}{2}g^{-1}v\cdot\partial
v^2+g^{\frac{1}{2}}K,\\
&&v\cdot\partial(\partial\cdot v)=-\frac{1}{2}g^{-2}(v\cdot\partial
v^2)^2-\frac{1}{2}g^{-1}v\cdot\partial(v\cdot\partial
v^2)+v\cdot\partial(g^{\frac{1}{2}}K),\\
&&\partial^2v^2=\partial_m\partial^m(v_n v^n) =\partial_m(v_n^m
v^n+v_n v^{mn}) =2\partial_m(v^{mn}
v_n)=2v\cdot\partial(\partial\cdot
v)+2v^{mn}v_{mn},\\
&&\begin{array}{ccl} v^m v^n\partial_{mn} v^2
&=&v^m_m[(\partial_n v^2)v^n]-v^m\partial_n v^2\partial_m
v^n=v^m\partial_m[(\partial_n v^2)v^n]-\partial_n
v^2v^m\partial^n v_m\\
&=&v\cdot\partial(v\cdot\partial v^2)-\frac{1}{2}\partial
v^2\cdot\partial v^2,
\end{array}
\end{array}\label{eq:videntities}
\end{equation}
where the first and the second equations are just $K$ expression, Eq.(\ref{eq:2ndorderA}) reduces to
\begin{equation}
\begin{array}{ccl}
&&g^{-\frac{1}{2}}g^{mn}\partial_m g^{\frac{1}{2}}\partial_n
g^{\frac{1}{2}}
-\frac{1}{2}\partial_n(g^{-\frac{1}{2}}g^{mn}\partial_m(g^{\frac{1}{2}})^2)\\
&=&\frac{3}{4}g^{-\frac{3}{2}}\partial v^2\cdot\partial
v^2+g^{-\frac{5}{2}}(v\cdot\partial
v^2)^2-\frac{1}{4}g^{-\frac{5}{2}}(v\cdot\partial v^2)^2
+\frac{1}{2}g^{-\frac{1}{2}}[2v\cdot\partial(\partial\cdot
v)+2v^{mn}v_{mn}]\\
&&+\frac{1}{2}g^{-\frac{3}{2}}[v\cdot\partial(v\cdot\partial
v^2)-\frac{1}{2}\partial v^2\cdot\partial v^2]+\frac{1}{2}g^{-1}K(v\cdot\partial v^2)\\
&=&g^{\frac{1}{2}}[\frac{1}{g}v^{mn}v_{mn}+\frac{1}{2g^2}\partial
v^2\cdot\partial v^2+\frac{1}{4g^3}(v\cdot\partial
v^2)^2]+\frac{1}{2}g^{-1}K(v\cdot\partial v^2)+g^{-\frac{1}{2}}v\cdot\partial(g^{\frac{1}{2}}K)\\
&=&g^{\frac{1}{2}}(K^a_bK^b_a)+\frac{1}{2}g^{-1}K(v\cdot\partial
v^2)+g^{-\frac{1}{2}}g^{-\frac{1}{2}}\frac{1}{2}v\cdot\partial(-v^2)K+g^{-\frac{1}{2}}g^{\frac{1}{2}}v\cdot\partial
K\\
&=&g^{\frac{1}{2}}(K^a_bK^b_a)+v\cdot\partial K,
\end{array}\label{eq:vdelK}
\end{equation}
where Eq.(\ref{eq:Kscalar}) has been used to convert the expression
to a compact form. When the equation of motion $K=0$ is applied, $R=K^a_bK^b_a-K^2=K^a_bK^b_a$ and $v\cdot\partial K=0$. Finally, Eq.(\ref{eq:vdelK})
becomes
\begin{equation}
g^{-\frac{1}{2}}g^{mn}\partial_m g^{\frac{1}{2}}\partial_n
g^{\frac{1}{2}}
-\frac{1}{2}\partial_n(g^{-\frac{1}{2}}g^{mn}\partial_m(g^{\frac{1}{2}})^2)=g^{\frac{1}{2}}R.\label{eq:gRA}
\end{equation}

\begin{acknowledgments}
The author is very grateful of many helps to Prof. T.E. Clark, Prof. T. ter Veldhuis, Prof. M. Kruczenski, Prof. S.T. Love, Dr. C. Xiong and Dr. A. Tirziu. Especially, he appreciates that Prof. T.E. Clark provided this project and reviewed the drafts with a great effort. The main necessary knowledge for the research came from discussions with him. Also, it was crucial that Prof. T. ter Veldhuis and Prof. M. Kruczenski explained to him the basic concepts through detailed calculations. Prof. S.T. Love, Dr. C. Xiong and Dr. A. Tirziu gave him a lot of valuable advice on the paper.
\end{acknowledgments}

\end{document}